\documentclass[twocolumn,english,aps,prb,showpacs,superscriptaddress,floats,amsmath,amssymb,floatfix]{revtex4}
\usepackage[latin9]{inputenc}
\usepackage{color}
\usepackage{amsmath}
\usepackage{graphicx}
\usepackage{graphics}
\usepackage[pdf]{pstricks}
\usepackage{pdftexcmds}
\usepackage{ifpdf}
\usepackage{amssymb}
\usepackage{esint}
\usepackage{multirow}
\usepackage{babel}

\makeatletter
\@ifundefined{textcolor}{}
{%
 \definecolor{BLACK}{gray}{0}
 \definecolor{WHITE}{gray}{1}
 \definecolor{RED}{rgb}{1,0,0}
 \definecolor{GREEN}{rgb}{0,1,0}
 \definecolor{BLUE}{rgb}{0,0,1}
 \definecolor{CYAN}{cmyk}{1,0,0,0}
 \definecolor{MAGENTA}{cmyk}{0,1,0,0}
 \definecolor{YELLOW}{cmyk}{0,0,1,0}
 }

\usepackage{array}
\usepackage{babel}\@ifundefined{definecolor}
{\usepackage{color}}{}

\newcommand{\beq}{\begin{equation}}
\newcommand{\eeq}{\end{equation}}

\newcommand{\bea}{\begin{eqnarray}}
\newcommand{\eea}{\end{eqnarray}}

\newcommand{\Sp}{\mathbf{S}}
\newcommand{\nn}{\nonumber}

\allowdisplaybreaks

\makeatother

\begin{document}

\title{Magnetization plateaux and jumps in a frustrated four-leg spin tube under a magnetic field}

\author{F.A. G\'omez Albarrac\'in}
\affiliation{IFLP-CONICET. Departamento de F\'{i}sica, Universidad Nacional de La Plata, C.C.
67, 1900 La Plata, Argentina}

\author{M.~Arlego}
\affiliation{IFLP-CONICET. Departamento de F\'{i}sica, Universidad Nacional de La Plata, C.C.
67, 1900 La Plata, Argentina}

\author{H.D.~Rosales }
\affiliation{IFLP-CONICET. Departamento de F\'{i}sica, Universidad Nacional de La Plata, C.C.
67, 1900 La Plata, Argentina}

\begin{abstract}
We study the ground state phase diagram of a frustrated spin-1/2 four-leg spin tube in an external magnetic field. We explore the parameter space of this model in the regime of all-antiferromagnetic exchange couplings by means of  three different approaches: analysis of low-energy effective Hamiltonian (LEH), a Hartree variational approach (HVA) and density matrix renormalization group (DMRG) for finite clusters. We find that in the limit of weakly interacting plaquettes, low-energy singlet, triplet and quintuplet  states play an important role in the formation of fractional magnetization plateaux. We study the transition regions numerically and analytically, and find that they are described, at first order in a strong- coupling expansion, by an XXZ spin-1/2 chain in a magnetic field; the second-order terms give corrections to the XXZ model. All techniques provide consistent results which allow us to predict the existence of fractional plateaux in an important region in the space of parameters of the model.
\end{abstract}


\maketitle
\section{Introduction}
\label{sec-intro}

Frustrated spin systems have been continuously explored in the last years driven by the role of frustration to induce unconventional magnetic orders or even disorder, including spin-liquid states and exotic excitations \cite{Diep-book, Lacroix-book}.
In particular, quasi one-dimensional spin systems, comprising chain, ladder and more involved magnetic structures are an active field of research thriving on a constant feedback between material synthesis, experimental investigations and theoretical predictions \cite{Dagotto1996a,Lemmens2003,Batchelor2007}.

Typically when these systems are placed in a magnetic field a richer behavior emerges ranging from the existence of fractional magnetization plateaux or the Bose-Einstein condensation of magnons to the possible existence of the spin-equivalent of a supersolid phase. Of particular interest are quasi one-dimensional systems as ladders and tubes, because they constitute an interesting and non trivial step from 1D to 2D.

As representative of geometrically frustrated homogeneous spin chains, one can consider the antiferromagnetic spin-1/2 zig-zag chain for which compounds such as CuGeO$_3$ \cite{CuGeO3}, LiV$_2$O$_5$ \cite{LiV2O5} or SrCuO$_2$ \cite{SrCuO2} are almost ideal prototypes and spin tube compounds with an odd number $N$ of sites per unit cell, such as {[}(CuCl$_{2}$tachH)$_{3}$Cl{]}Cl$_{2}$ \cite{Schnack2004} and CsCrF$_{4}$ \cite{Manaka2009} with $N=3$,
and Na$_{2}$V$_{3}$O$_{7}$ \cite{Millet1999} with $N=9$. Note that spin tubes with an odd number of legs and only nearest neighbor antiferromagnetic (AFM) exchange are \emph{geometrically} frustrated.

Recently, Cu$_{2}$Cl$_{4}$$\cdot$D$_{8}$C$_{4}$SO$_{2}$ has been established as a new spin-1/2 tube with an even number of legs \cite{Garlea2008a}, namely $N=4$. Tubes with $N=4$ and only nearest
neighbor AFM exchange are \emph{not} frustrated. However, substantial next-nearest neighbor AFM exchange, diagonally coupling adjacent legs, has been claimed for Cu$_{2}$Cl$_{4}$ $\cdot$D$_{8}$C$_{4}$SO$_{2}$,
rendering also this ladder system frustrated.

\begin{figure}[tb]
\begin{centering}
\includegraphics[width=0.8\columnwidth]{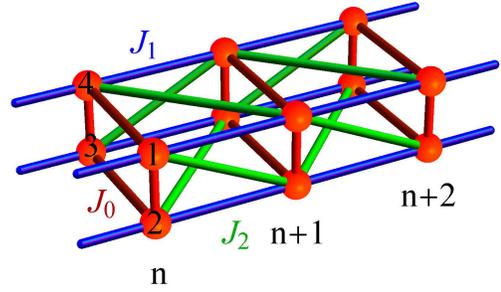}
\par\end{centering}
\caption{\label{fig1}(Color online). Frustrated four-spin tube. Solid spheres represent
spin-$1/2$ moments and the labels $1,2,3,4$ indicate the four sites of each unit cell. Plaquettes (bold red lines) are coupled by nearest ($J_{1}$) and next nearest ($J_{2}$) antiferromagnetic exchange,
blue and green lines, respectively. On-plaquette coupling is $J_{0}$.}
\label{fig:model}
\end{figure}

Motivated by this, in this paper we study the \emph{geometrically} frustrated  four-leg spin tube (FFST) model that has been introduced in ref. \cite{Arlego2011,Arlego2012}, in presence of a magnetic field. The Hamiltonian is given by
\beq
\mathcal{H}=\mathcal{H}_{\text{plaq}}+\mathcal{H}_{\text{int}},
\label{eq:Ham}
\eeq
where $\mathcal{H}_{\text{plaq}}$ contains the interactions between spins in each plaquette plus the Zeeman term,
\beq
\mathcal{H}_{\text{plaq}}=J_0\sum_{n,a}\Sp_{n,a}\cdot\Sp_{n,a+1}-h\sum_{n,a}S^{z}_{n,a},
\label{eq:Ham-plaq}
\eeq
and $\mathcal{H}_{\text{int}}$  contains the Heisenberg interactions between adjacent plaquettes
\beq
\mathcal{H}_{\text{int}}=J_1\sum_{n,a}\Sp_{n,a}\cdot\Sp_{n+1,a}+J_2\sum_{n,a}\Sp_{n,a}\cdot\Sp_{n+1,a+1},
\label{eq:Ham-int}
\eeq
with the lattice structure and exchange antiferromagnetic couplings $J_{0,1,2}$ as shown in Fig.\ \ref{fig:model}. Here $a=1,...,4$ (resp. $n=1,...,N$) is a site (resp. plaquette) index, $J_0$ is the coupling on each plaquette, $J_{1,2}$ the couplings along the chains and the site $(n, 5)$ is identified with the site $(n, 1)$. Note that the FFST model can be mapped onto an identical one with exchanged $J_1 \leftrightarrow  J_2$ by a $\pi/2$ twist of the plaquettes around the tube.

For $J_{1,2}\ll J_{0}$, the quantum properties of the FFST can be understood in terms of weakly coupled four-spin plaquettes. In ref. [\onlinecite{Arlego2011}]  a series expansion analysis of the one- and two-particle excitations has been carried out for the case of zero magnetic field in this restricted parameter regime. In [\onlinecite{Arlego2012}] by a combined analysis from a variety of complementary methods, the complete parameter space of the FFST has been explored. However, a study  of the phases of the FFST in the presence of a magnetic field has not been done.

In this paper we pay particular attention to the behavior of the model in the limit of weakly coupled plaquettes where it is possible to obtain an effective description in terms of degenerate perturbation theory. We find that the effect of frustrating interactions leads to the appearance of additional fractional magnetization plateaux, which have already been shown to exist in several frustrated quasi-1D systems\cite{Mila2006,LEH}.  In a combined analysis using perturbative methods, variational approach  and the density matrix renormalization group (DMRG), we make quantitative predictions for the existence,  the position and the sizes of these plateaux induced by frustration.

The paper is structured as follows: In Sec. \ref{sec-LEEM} we derive the low energy effective hamiltonian of the model given by Eq.(\ref{eq:Ham}). After that, by means of Bethe-Ansatz analysis in Sec. \ref{sec-Bethe-Ansatz}, the low-energy dispersion calculation near the ends of plateaux in Sec. \ref{sec-MagnonDisper} and a variational approach in Sec. \ref{sec-VA}, we show that for certain values of the frustrating parameters, the ground state can spontaneously break translation invariance symmetry leading to  additional plateaux at intermediate values of the magnetization.
In Sec. \ref{sec-Discussion-Conclusions} we present an analysis of the phase diagram obtained in the range of parameters considered, and we finish the paper in Sec. \ref{sec-Conclusions} with a summary of the main results obtained in the work, its scope, and possible extensions for future studies.

Let us finally mention that numerical work  is not specifically presented in a given Section, but rather throughout the text, to allow a closer comparison between low energy effective models predictions and finite size numerical calculations on the spin tube model using DMRG technique.

\section{Low Energy Effective Models}
\label{sec-LEEM}

The physics of the model given by Eq. (\ref{eq:Ham}) is controlled by two factors, the level of frustration of the Heisenberg exchange and the magnetic field. In the limit  $J_{1,2} = 0$, the system consists of independent plaquettes. The Hilbert space of each plaquette contains sixteen states which fall into  two spin-0 singlets ($|s^{(1)}\rangle$, $|s^{(2)}\rangle$),
nine spin-1 triplets ($|t^{(1)}_{i}\rangle$, $|t^{(2)}_{i}\rangle$ and  $|t^{(3)}_{i}\rangle$ with $i=1,...,3$) and five spin-2 quintuplet $|q_{i}\rangle$ (with $i=1,...,5$). These states are listed in table \ref{table-plaquette-states} where $|S\rangle_{ab}$ is a singlet state between sites $a$ and $b$ defined as $|S\rangle_{ab}=(|+,-\rangle_{ab}-|-,+\rangle_{ab})/\sqrt{2}$, $|t\rangle^{0}_{ab}=(|+,-\rangle_{ab}+|-,+\rangle_{ab})/\sqrt{2}$ and $|t\rangle^{\pm}_{ab}=|\pm,\pm\rangle_{ab}$.

\begin{table}
\begin{center}

\begin{tabular}{|c|c|}
\hline
\multicolumn{2}{ |c| }{Plaquette states} \\
\hline
$|s^{(1)}\rangle$ &$\frac{|S\rangle_{14}\,|S\rangle_{23}-|S\rangle_{12}\,|S\rangle_{34}}{\sqrt{3}}$\\
\hline
$|s^{(2)}\rangle$&$\frac{|t\rangle^{-}_{12}|t\rangle^{+}_{34}+|t\rangle^{+}_{12}|t\rangle^{-}_{34}-|t\rangle^{-}_{14}|t\rangle^{+}_{23}-|t\rangle^{+}_{14}|t\rangle^{-}_{23}}{2}$\\
\hline
$|t^{(1)}_1\rangle$&$\frac{|S\rangle_{12}|t\rangle^{+}_{34} +|t\rangle^{+}_{12}|S\rangle_{34}}{\sqrt{2}}$\\
\hline
$|t^{(1)}_2\rangle$&$\frac{|S\rangle_{12}|t\rangle^{0}_{34} + |t\rangle^{0}_{12}|S\rangle_{34}}{2}$\\
\hline
$|t^{(1)}_3\rangle$&$\frac{|S\rangle_{12}|t\rangle^{-}_{34} +|t\rangle^{-}_{12}|S\rangle_{34}}{\sqrt{2}}$\\
\hline
$|t^{(2)}_1\rangle$&$|S\rangle_{13}|t\rangle^{+}_{24}$\\
\hline
$|t^{(2)}_2\rangle$&$|t\rangle^{-}_{12}|t\rangle^{+}_{34}-|t\rangle^{+}_{12}|t\rangle^{-}_{34}$\\
\hline
$|t^{(2)}_3\rangle$&$|S\rangle_{24}|t\rangle^{-}_{13}$\\
\hline
$|t^{(3)}_1\rangle$&$|S\rangle_{24}|t\rangle^{+}_{13}$\\
\hline
$|t^{(3)}_2\rangle$&$|t\rangle^{-}_{14}|t\rangle^{+}_{23}-|t\rangle^{+}_{14}|t\rangle^{-}_{23}$\\
\hline
$|t^{(3)}_3\rangle$&$|S\rangle_{13}|t\rangle^{-}_{24}$\\
\hline
$|q_1\rangle$&$ |t\rangle^{+}_{12}|t\rangle^{+}_{24}$\\
\hline
$|q_2\rangle$&$ \frac{|t\rangle^{0}_{12}|t\rangle^{+}_{34}+|t\rangle^{+}_{12}|t\rangle^{0}_{34}}{\sqrt{2}}$\\
\hline
$|q_3\rangle$&$ \frac{|t\rangle^{0}_{12}|t\rangle^{0}_{34}+|t\rangle^{0}_{14}|t\rangle^{0}_{23}+|t\rangle^{0}_{13}|t\rangle^{0}_{24}}{\sqrt{6}}$\\
\hline
$|q_4\rangle$&$ \frac{|t\rangle^{0}_{12}|t\rangle^{-}_{34}+|t\rangle^{-}_{12}|t\rangle^{0}_{34}}{\sqrt{2}}$\\
\hline
$|q_5\rangle$&$ |t\rangle^{-}_{12}|t\rangle^{-}_{24}$\\
\hline
\end{tabular}
\end{center}
\caption{Plaquette states of the Hamiltonian $\mathcal{H}_{\text{plaq}}$ in Eq. (\ref{eq:Ham-plaq}).}
\label{table-plaquette-states}
\end{table}

\begin{figure}[tb]
\begin{centering}
\includegraphics[width=1\columnwidth]{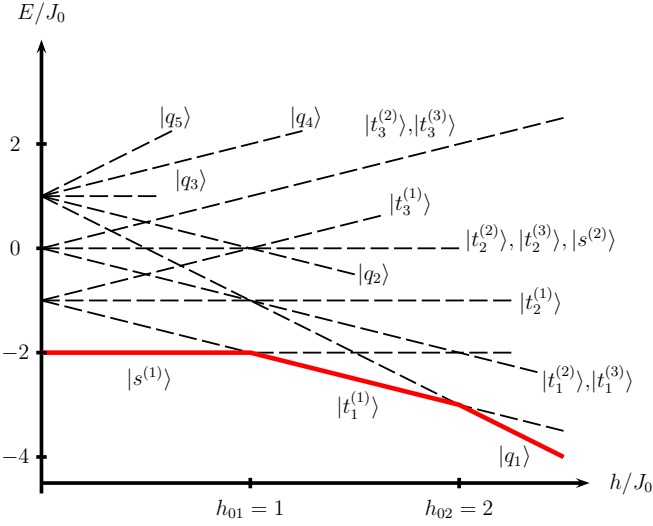}
\par
\end{centering}
\caption{(Color online). Level crossings of a single plaquette of $S=1/2$. The full red line denotes the ground state at different values of the magnetic field. There are two values of $h$ where the ground state is degenerate, for $h_{01}=J_0$ and $h_{02}=2\,J_0$, where the ground state changes from singlet to triplet and from triplet to quintuplet, respectively. Note that at these magnetic fields the separation between
the energy levels is $J_0$, defining the energy scale for a perturbative treatment in the weakly coupled four-spin plaquettes limit.
}
\label{fig:Epla-vs-H}
\end{figure}

When an external magnetic field is switched on the degeneracy in the different multiplets is lifted. As shown in Fig.\ref{fig:Epla-vs-H},  there are two ground state level crossings at two values of the magnetic field, $h_{01}=J_0$  and $h_{02}=2\,J_0$. At these values, the ground state is degenerate. For $h<h_{01}=J_0$, the ground state is $|s^{(1)}\rangle$; for $h_{01}<h<h_{02}=2\,J_0$ the ground state is $|t^{(1)}_1\rangle$ while for $h>h_{02}$ the ground state is $|q_1\rangle$.

We will now discuss the low-energy effective Hamiltonian (LEH) approach  used to study  the properties of the spin tube given by Eq. (\ref{eq:Ham}). There are two possible limits which may be  considered.

One limit is the case $J_0/J_{1,2} \rightarrow 0$, which corresponds to weakly interacting chains, that can be analyzed by means of bosonization  and conformal field theory;
this has been done in detail by other authors[\onlinecite{cabratotsuka}]. The other limiting case, that we will consider here, is the strong-coupling limit $J_{1,2}/J_0 \ll 1$ which corresponds to almost decoupled plaquettes,
and where the inter-plaquette couplings can be treated perturbatively.

We derive the LEH as follows: we first set the inter-plaquette couplings $J_{1,2}=0$ and select the states of a single plaquette which are degenerate in energy in  the presence of a magnetic field. As mentioned above, there are two such values of the magnetic field in this case. We will consider each such value of $h_0$  separately. The degenerate plaquette states will constitute our low-energy states.
Next, using the Hamiltonian of the total system as $\mathcal{H}= \mathcal{H}_{\text{plaq}}'+\mathcal{H}_{\text{int}}'$, where $\mathcal{H}_{\text{plaq}}'=\mathcal{H}_{\text{plaq}}-h_0\sum_{j,a}S^{z}_{j,a}$ and $\mathcal{H}_{\text{int}}'=\mathcal{H}_{\text{int}}-(h-h_0)\sum_{j,a}S^{z}_{j,a}$ contains the small interactions $J_{1,2}$ and the residual magnetic field $h - h_0$ which are both assumed to be much smaller than $J_0$\textcolor[rgb]{1.00,0.00,0.00}{.}
Let us now denote the degenerate and low-energy states of the  system as $v_i$ and the high-energy states as $w_{\alpha}$. The low-energy states all have
energy
$\epsilon_0$, while the high-energy states have energies $\epsilon_{\alpha}$ according to the exactly solvable Hamiltonian $\mathcal{H}_{\text{plaq}}'$. With this we construct an effective Hamiltonian\cite{LEH}
\beq
\mathcal{H}_{\text{eff}}=\mathcal{H}^{(1)}_{\text{eff}}+\mathcal{H}^{(2)}_{\text{eff}}+...
\label{eq:Heff}
\eeq
where $\mathcal{H}_{\text{eff}}^{(i)}$ is the $ith$ order of the perturbation expansion. The first-order term is
\beq
\mathcal{H}_{\text{eff}}^{(1)} = \sum_{ij} \vert v_i \rangle \langle v_i \vert \mathcal{H}_{\text{int}}' \vert v_j  \rangle ~\langle v_j \vert .
\label{eq:Heff1}
\eeq
The second-order LEH is given by
\beq
\mathcal{H}_{\text{eff}}^{(2)} = \sum_{ij} ~\sum_{\alpha} ~\vert v_i \rangle ~\frac{\langle
v_i \vert  \mathcal{H}_{\text{int}}' \vert w_{\alpha} \rangle ~\langle w_{\alpha} \vert \mathcal{H}_{\text{int}}' \vert
v_j \rangle}{\epsilon_0 -\epsilon_{\alpha}} ~\langle v_j \vert ~.
\label{eq:Heff2}
\eeq

Finally, we introduce pseudo-spin operators representing the states (around each  magnetic field ($h_{01}$ and $h_{02}$)) of each plaquette to rewrite the effective Hamiltonian in a more transparent form amenable for further analysis. The  effective Hamiltonian up to second order obtained according to the procedure described above is:

\noindent
(i)\,For $h_{01}=J_0$, the two states to be considered are  $|s^{(1)}\rangle$ and $|t^{(1)}_1\rangle$ with energies  $-2\,J_0$ and $-J_0-h$ respectively.
The corresponding operators are:
\bea
S^{z}_n=\frac{1}{2}\left(|t^{(1)}_1\rangle\langle t^{(1)}_1|-|s^{(1)}\rangle\langle s^{(1)}|\right),\nn\\
S^{+}_n=|s^{(1)}\rangle\langle t^{(1)}_1|,\quad S^{-}_n=|t^{(1)}_1\rangle\langle s^{(1)}|
\eea

\noindent
(ii)\,In the case of $h_{02}=2\,J_0$ the relevant two states of the plaquette to take into account are: $|t^{(1)}_1\rangle$ and $|q_1\rangle$ with energies $-J_0-h$ and $J_0-2\,h$ respectively,
with operators:

\bea
S^{z}_n=\frac{1}{2}\left(|q_1\rangle\langle q_1|-|t^{(1)}_1\rangle\langle t^{(1)}_1|\right),\nn\\
S^{+}_n=|t^{(1)}_1\rangle\langle q_1|,\quad S^{-}_n=|q_1\rangle\langle t^{(1)}_1|
\eea
In both cases, the form of the effective Hamiltonian is the same but the values of the effective couplings change. Therefore, Eq. (\ref{eq:Heff}) becomes
\bea
\mathcal{H}_{\text{eff}}&=&\epsilon_0+\sum_{j=1}^{N}J_{xy}(S^{x}_{j}S^{x}_{j+1}+S^{y}_{j}S^{y}_{j+1})+J_{zz}\,S^{z}_{j}S^{z}_{j+1}\nn\\
&&+\sum_{j=1}^{N}K_{xy}(S^{x}_{j}S^{x}_{j+2}+S^{y}_{j}S^{y}_{j+2})-h^{\text{eff}}\sum_{j=1}^{N}S^{z}_{j}\nn\\
&&+\sum_{j=1}^{N}J_{xyz}(S^{x}_{j}S^{x}_{j+2}+S^{y}_{j}S^{y}_{j+2})S^{z}_{j+1}
\label{eq:Heff_LEH}
\eea
where around the first magnetic field $h_{01}$ effective couplings and field are given by
\bea
\frac{\epsilon_0}{N}&=&-2J_0+\frac{J_1+J_2}{16}-\nn\\
&&\frac{31}{6912}\frac{109(J_1^2+J_2^2)-194\,J_1\,J_2}{J_0}\nn\\
h^{\text{eff}}&=&h-J_0-\frac{J_1+J_2}{4}+\frac{11(J_1^2+J_2^2)-14J_1J_2}{192}\nn\\
J_{xy}&=&\frac{4(J_1-J_2)}{3}+\frac{(J_1^2-J_2^2)}{9\,J_0}\nn\\
J_{zz}&=&\frac{(J_1+J_2)}{4}-\frac{1715(J_1^2+J_2^2)-3454 J_1J_2}{1728\,J_0}\nn\\
K_{xy}&=&-\frac{31(J_1-J_2)^2}{54\,J_0}\nn\\
J_{xyz}&=&-\frac{7(J_1-J_2)^2}{27\,J_0}
\label{eq:LEH-constants-1}
\eea
whereas around the second one $h_{02}$ they are given by

\bea
\frac{\epsilon_0}{N}&=&J_0+\frac{9(J_1+J_2)}{16}-\nn\\
&&\frac{1}{256}\frac{49(J_1^2+J_2^2)-90\,J_1\,J_2}{J_0}\nn\\
h^{\text{eff}}&=&h-2\,J_0-\frac{3(J_1+J_2)}{4}-\frac{49(J_1^2+J_2^2)-90J_1J_2}{64\,J_0}\nn\\
J_{xy}&=&J_1-J_2\nn\\
J_{zz}&=&\frac{J_1+J_2}{4}-\frac{49(J_1^2+J_2^2)-90J_1J_2}{64\,J_0}\nn\\
K_{xy}&=&-\frac{11(J_1-J_2)^2}{32\,J_0}\nn\\
J_{xyz}&=&\frac{11(J_1-J_2)^2}{16\,J_0}
\label{eq:LEH-constants-2}
\eea

In both cases, the first order effective Hamiltonian becomes an XXZ model, where the Bethe-Ansatz solution can be used to obtain information about the system. In the special case where both inter-plaquette couplings are equal $J_1=J_2$, the effective couplings $J_{xy}$, $K_{xy}$ and $J_{xyz}$ become $0$, and the model reduces to an effective Ising Hamiltonian. Furthermore, from the expressions (\ref{eq:LEH-constants-1}) and (\ref{eq:LEH-constants-2}) we see that in both cases the constants $K_{xy}$ and $J_{xyz}$ are order $(J_1-J_2)^2$ while $J_{xy}$, $J_{zz}$ and $h^{\text{eff}}$ contains  a term proportional to $(J_1-J_2)$. So, we expect that close to the line $J_1=J_2$ the second order corrections will be more important in the effective XXZ model that in  $K_{xy}$ and $J_{xyz}$. This will be discussed in following sections.

\begin{figure}[tb]
\begin{centering}
\includegraphics[width=0.95\columnwidth]{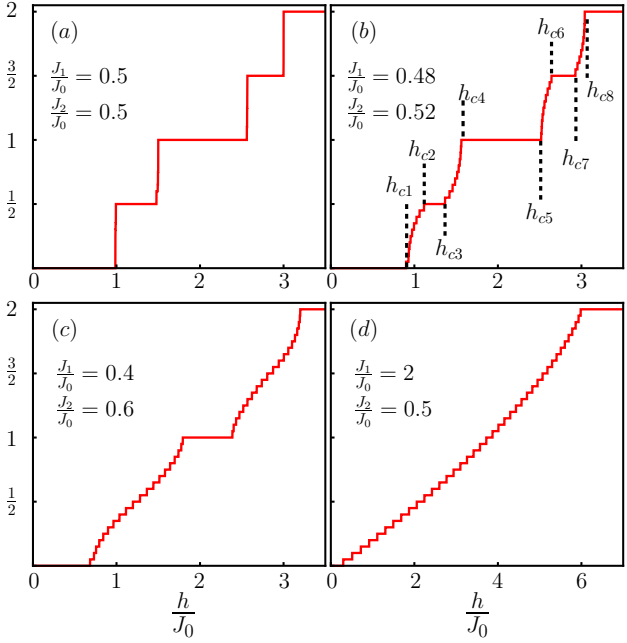}
\end{centering}
\caption{(Color online). Magnetization by plaquette vs magnetic field obtained by DMRG simulations for with open boundary conditions (OBC) for  $N=4\times L$ with $L=20$.}
\label{fig:fields_hc}
\end{figure}
%

\subsection{Bethe-Ansatz solution of effective model}
\label{sec-Bethe-Ansatz}

As mentioned previously, the effective models obtained around $h_{01}$ and $h_{02}$ reduce to an XXZ effective spin-1/2 model at first order, $O(J_1, J_2)$, since that in both cases $K_{xy}$ and $J_{xyz}$  are second order terms, i.e.

\bea
\mathcal{H}_{\text{eff}}&=&\sum_{j=1}^{N}J_{xy}(S^{x}_{j}S^{x}_{j+1}+S^{y}_{j}S^{y}_{j+1})+J_{zz}\,S^{z}_{j}S^{z}_{j+1}\nn\\
&&-h^{\text{eff}}\sum_{j=1}^{N}S^{z}_{j}.\nn\\
\label{eq:Heff-XXZ}
\eea

As it is known this model has exact solution via Bethe-Ansatz, which will allow us to predict main
physical features of the tube model (at least in the range of weakly coupled plaquettes)\cite{BetheAns}. To this end let us first briefly review the main characteristics of XXZ chain. In absence of a magnetic field $h^{\text{eff}}=0$ and for $-1 \leq \Delta \leq 1$ ($\Delta=J_{zz}/|J_{xy}|$) the system is in a gapless Luttinger liquid phase. For $ \Delta < -1$ the ground state is ferromagnetic with a gap to effective spin-1 magnon excitations. On the other hand for $ \Delta >1$, the system exhibits a N\'eel ordered phase with a gap to effective spin-1/2 domain-wall \emph{spinon} excitations. Elementary magnon (spinon) excitation condense at the boundary $ \Delta = -1 (1)$.

\noindent In presence of a field $h^{\text{eff}}$, in the plane $\Delta -  h^{\text{eff}}, $ there are two critical lines $h^{\text{eff}}_{\text{L-F}}$ and $h^{\text{eff}}_{\text{L-N}}$ confining Luttinger liquid phase between ferromagnetic and N\'eel phases, which are given by\cite{BetheAns}

\begin{eqnarray}
  \pm h^{\text{eff}}_{\text{L-F}} &=& |J_{xy}|+J_{zz},\nonumber \\
  \pm h^{\text{eff}}_{\text{L-N}}&=& |J_{xy}|
  \sinh g \sum_{n=-\infty}^\infty{\frac{(-1)^n}{\cosh (n g)}}, \nonumber \\
  \label{eq:Hc-BA}
\end{eqnarray}
where $\cosh g = \Delta$. The gapped phases translates into plateaux in magnetization curves, at M=0 for N\'eel and trivially at $M=\pm \frac{1}{2}$ (normalized per site) for ferromagnetic phase. On the other hand, in the gapless Luttinger liquid phase magnetization increases continuously with the applied field.

This simple structure of magnetization curve, connecting a central integer plateau with half-integer plateaux at each side, via Luttinger liquid phases, is the  main feature that describes
qualitatively the magnetization curve of the tube model in the strong coupling regime.
\noindent Specifically for the field sector around $h_{01}$, the effective $S_z=-\frac{1}{2}(+\frac{1}{2})$ represents the plaquette $S_z=0(1)$, corresponding to the singlet(triplet) involved in the low field crossing. Therefore the plateaux in the curve of magnetization (per site) vs $h^{\text{eff}}$ in the effective model at $M=-\frac{1}{2}$, $0$ and $\frac{1}{2}$ translates into plateaux at $M=0$, $\frac{1}{2}$ and $1$ in the curve of magnetization (per plaquette) vs $h\geq0$. The same idea applies to the field sector around $h_{02}$, where effective $S_z=-\frac{1}{2}(+\frac{1}{2})$ represents plaquette $S_z=1(2)$, corresponding to triplet(quintet) at the high field crossing. In this case, plateaux at $M=0$, $\frac{1}{2}$ and $1$ in effective model translates into $M=1$, $\frac{3}{2}$ and $2$ in plaquette model.

The critical lines, limiting integer plateaux, Luttinger liquid and half-integer plateaux phases, are determined by solving numerically the Eqs.(\ref{eq:Hc-BA}) corresponding to  the effective models
around $h_{01}$ and $h_{02}$. This calculation, together with the comparison with the other techniques (see Fig. \ref{fig:h_vs_beta_eff12}) is discussed in Section \ref{sec-Discussion-Conclusions}. \\

Along the line of maximum frustration $J_1=J_2$, the system reduces to an effective Ising model, with a single transition between N\'eel and ferromagnetic phase, and absence of Luttinger liquid phase, since $J_{xy}=0$ for both  Eqs.(\ref{eq:LEH-constants-1}) and Eqs.(\ref{eq:LEH-constants-2}). This is reflected in the magnetization curve by means of a stepwise structure with jumps between plateaux at integer and half-integer values of $M$ per plaquette, corresponding to ferromagnetic and N\'eel phases, respectively.\\

By applying the previous condition $J_{xy}=0$ to first of Eq. (\ref{eq:Hc-BA}) and using the corresponding $J_{zz}$ from the Eqs.(\ref{eq:LEH-constants-1}) and Eqs.(\ref{eq:LEH-constants-2}), respectively, we obtain, at $O(J_1,J_2)$ the following expressions for the critical lines, in units of $J_0$(see Fig. \ref{fig:fields_hc}b)
\begin{equation}
    h_{c1}=1,\quad h_{c3}=1+J_1,\quad h_{c5}=2+J_1,\quad h_{c7}=2+2 J_1,\\
    \label{eq:Hc-BA-diag}
\end{equation}
and $h_{c2}= h_{c1}$, $h_{c4}= h_{c3}$, $h_{c6}= h_{c5}$, and $h_{c8}= h_{c7}$.

One aspect which is important to recall is the role of frustration on the plateaux structure
of the tube model. There is a crucial difference between both types of plateaux regarding the influence of frustration. Integer-type of plateaux are inherent of each plaquette, i.e. they exist independently
of the inter-plaquette coupling (although they are renormalized by them). On the other hand, half-integer plateaux are induced by frustration and are widest along $J_1=J_2$. Around that line, frustration-induced
plateaux start to narrow (as well as integer plateaux), leaving space to a growing Luttinger liquid phase which is the only one that survives in the limit of decoupled chains.

To check low energy results obtained by Bethe-Ansatz we have performed extensive DMRG computations[\onlinecite{alps}] on the model Hamiltonian given by Eqs.(\ref{eq:Ham}-\ref{eq:Ham-int}). We calculate the magnetization per plaquette  $M=S^z_{total}/L$, being $L$ the number of plaquettes in the spin-tube. In our DMRG calculations we have employed  periodic (PBC) boundary conditions, and keeping up to 500 states, which has shown to be enough to ensure the expected accuracy.

In figure \ref{fig:fields_hc} we show numerical DMRG results for magnetization curves for spin-tubes composed by $L=20$ plaquettes and PBC. The $J_{1,2}$ (in units of $J_0$) values have been selected in order to illustrate the emergence of different plateaux structures discussed previously. The left-upper panel shows the predicted Ising-like behavior along the line of maximum frustration for the case $J_1/J_0=J_2/J_0=0.5$, with plateaux boundaries satisfying Eqs.(\ref{eq:Hc-BA-diag}). Small deviations from $J_1=J_2$ line induce  a reduction of half-integer plateaux and the transitions from jumps to smooth curves between plateaux, characteristic of Luttinger liquid phases. This behavior is shown in the right-upper of Fig. \ref{fig:fields_hc}, for the case $J_1/J_0=0.48$ and $J_2/J_0=0.52$.

The presence of half-integer plateaux is very sensitive to frustration. This is illustrated in left-lower panel of Fig. \ref{fig:fields_hc}, which shows that already for $J_1/J_0=0.4$ and $J_2/J_0=0.6$ half-integer plateaux are not present, remaining only a integer plateaux structure connected by Luttinger liquid phases.
Finally, far from the $J_1=J_2$ line and near to $J_1/J_0=0$ or $J_2/J_0=0$ lines, the magnetization shows only a gapless Luttinger Liquid phase, which is a signature that the system is adiabatically connected with the limit of decoupled spin-1/2 chains. This behavior is shown in the right-lower panel of Fig. \ref{fig:fields_hc}, for the values $J_1/J_0=2$ and $J_2/J_0=0.5$.

\subsection{Magnon and Spinon dispersion}
\label{sec-MagnonDisper}

%
\begin{figure}[tb]
\begin{centering}
\includegraphics[width=0.95\columnwidth]{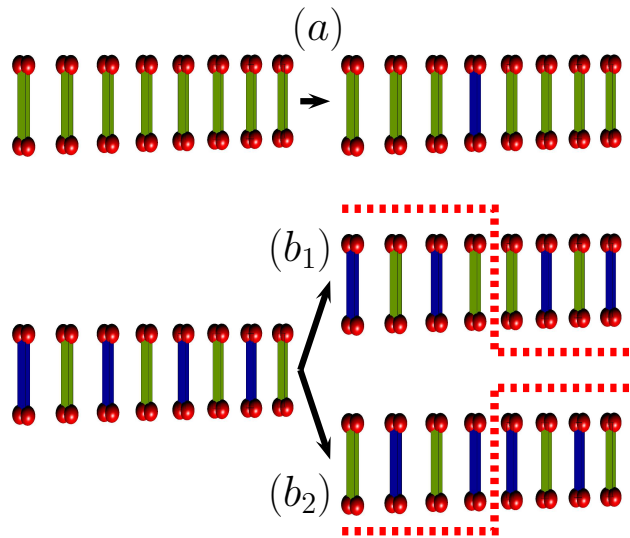}
\end{centering}
\caption{(Color online). A sketch of the low energy excitations for (a) integer plateaux: all of the plaquettes are
in one state and the excitations out of this state correspond to magnons carrying spin $\Delta S^z=\pm 1$;
(b) half-integer plateau: Elementary excitations are domain walls (two plaquettes in the same state) carrying
$\Delta S^z = +1/2$ for the low-field end of the plateau ($b_1$) and  $\Delta S^z = -1/2$ for the low-field beginning of the plateau ($b_2$).}
\label{fig:GS_and_First_Exc}
\end{figure}

We will now use the effective Hamiltonian given by Eq. (\ref{eq:Heff_LEH}) to compute the values of the critical fields $h_{c_1}...h_{c_8}$ (see Fig. \ref{fig:fields_hc}b) by means of the analysis of  the gap in the spectrum of low-energy excitations at the ends of the plateaux\cite{LEH,Mila2006}. To start we compute the critical field $h_{c_4}$ at the beginning of the $M=1$ plateau where the state with all plaquettes equal to $|t^{(1)}_1\rangle$ becomes the ground state. The elementary excitations correspond to a superposition of individual singlets  $|s^{(1)}\rangle$ carrying spin $\Delta S^z=-1$ in a background of triplets (see Fig. \ref{fig:GS_and_First_Exc}$a$). To compute the field $h_{c_4}$, we compare the energy $\epsilon_0$ of the state with all plaquettes in $|t^{(1)}_1\rangle$ with the minimum energy of a spin-wave state in which one plaquette is in $|s^{(1)}\rangle$ and all the other plaquettes in $|t^{(1)}_1\rangle$. A spin wave with momentum $k$ is given by\cite{LEH}

\beq
|k\rangle=\frac{1}{\sqrt{L}}\sum_n\,e^{i\,kn}\,|s^{(1)}\rangle_n
\label{eq:spin_wave_hc4}
\eeq
where $|s^{(1)}\rangle_n$ denotes a state with a singlet $|s^{(1)}\rangle$ on plaquette at site $n$,
and triplets $|t^{(1)}_1\rangle$ on the other sites. The spin-wave dispersion, i.e. $w(k)=\epsilon(k)-\epsilon_0$, is obtained by applying effective Hamiltonian (Eq. (\ref{eq:Heff_LEH})) to Eq. (\ref{eq:spin_wave_hc4}) i.e. $\mathcal{H}_{\text{eff}}|k\rangle =(w(k)+\epsilon_0)|k\rangle $, in this case we get

\bea
w_{4}(k)&=&C^{+}_0+C_1\,\cos(k)+C^{+}_2\cos(2k)
\label{eq:whc4}
\eea
with

\bea
C^{\pm}_0=\pm h^{\text{eff}}-J_{zz},\, C_1=J_{xy},\, C^{\pm}_{2}=K_{xy}\pm \frac{J_{xyz}}{2}
\label{eq:w0pmw1w2pm}
\eea
By setting $w_{4}(k^*)=0$ at $k^*$ where the spin wave dispersion $w_{4}(k)$ has a minimum, we obtain the  critical field $h_{c4}$ in terms of $J_0,J_1,J_2$. Similarly, we compute the fields $h_{c1},h_{c5}$ and $h_{c8}$, delimiting integer plateaux  by comparing the energy $\epsilon_0$ of the state with all plaquettes are in the state composed by $|s^{(1)}\rangle$, $|t^{(1)}_1\rangle$ and $|q_1\rangle$ on all plaquettes, respectively, with the minimum energy $\epsilon_{min}(k)$  of a spin wave in which one state is replaced by a $|t^{(1)}_1\rangle$,  $|q_1\rangle$ and $|t^{(1)}_1\rangle$ respectively. The spin-wave dispersions $w_{1}(k)$, $w_{5}(k)$, $w_{8}(k)$ that we obtain are

\bea
w_{1}(k)&=&C^{-}_0+C_1\,\cos(k)+C^{-}_2\cos(2\,k)\nn\\
w_{5}(k)&=&C^{-}_0+C_1\,\cos(k)+C^{-}_2\cos(2\,k)\nn\\
w_{8}(k)&=&C^{+}_0+C_1\,\cos(k)+C^{+}_2\cos(2\,k)
\label{eq:whc158}
\eea

Note that in the previous expressions the coefficients involved (see Eq. (\ref{eq:w0pmw1w2pm})) depend on the critical field considered. For $w_{1}(k)$ and $w_{4}(k)$ the coefficients $J_{xy}$, $J_{zz}$, $h^{\text{eff}}$,  $K_{xy}$  and $J_{xyz}$  are given by Eqs. (\ref{eq:LEH-constants-1}) whereas for $w_{5}(k)$ and $w_{8}(k)$ by Eqs. (\ref{eq:LEH-constants-2}).

In the case of the fractional plateaux at $M=1/2$ and $3/2$ the low-energy excitations (near the ends of the plateaux) are no longer magnons, but  are domain walls with spin $S^z = \pm 1/2$ (see figures  \ref{fig:GS_and_First_Exc}$b_1$ and \ref{fig:GS_and_First_Exc}$b_2$). The critical fields can be obtained by considering the dispersion of the corresponding elementary excitation on the adequate background, following a similar procedure that the employed for the case of integer-plateaux presented before. The spin wave dispersions for half-integer plateaux are

\bea
w_{2}(k)&=&\frac{J_{zz}}{2}+\frac{h^{\text{eff}}}{2}+J_{xy}\,\cos(2\,k)\nn\\
w_{3}(k)&=&\frac{J_{zz}}{2}-\frac{h^{\text{eff}}}{2}+J_{xy}\,\cos(2\,k)\nn\\
w_{6}(k)&=&\frac{J_{zz}}{2}+\frac{h^{\text{eff}}}{2}+J_{xy}\,\cos(2\,k)\nn\\
w_{7}(k)&=&\frac{J_{zz}}{2}-\frac{h^{\text{eff}}}{2}+J_{xy}\,\cos(2\,k)
\label{eq:whc2367}
\eea
where $h^{\text{eff}}$, $J_{zz}$ and $J_{xy}$ are given by Eqs. (\ref{eq:LEH-constants-1}) for $w_{2}(k)$ and $w_{3}(k)$, and by Eqs. (\ref{eq:LEH-constants-2}) for $w_{6}(k)$ and $w_{7}(k)$.

It is simple to see that along the line $J_1=J_2$ the dispersion relations are flat because the amplitudes of the cosines are canceled. This will be reflected in the step-wise structure of magnetization curve along that line.

The critical lines, limiting integer plateaux, Luttinger liquid and half-integer plateaux phases, obtained by analyzing previously calculated dispersions, together with a comparison with the other techniques (Fig. \ref{fig:h_vs_beta_eff12}) are presented in Section \ref{sec-Discussion-Conclusions}.


\subsection{Variational approach}
\label{sec-VA}
An alternative way to study the low energy properties of the spin tube in presence of a magnetic field is by means of a variational approach\cite{Varia1}. To this end we consider a Hartree variational function consisting of a linear combination of eigenstates of the plaquette per plaquette, i.e.
the wave function will be of the form\cite{Varia1}
\begin{equation}
|\psi\rangle=\prod^{L}_{n=1}\,|\nu_n\rangle
\label{eq:psi_VA}
\end{equation}
where $|\nu_n\rangle=\sum_{i=1}^{Ne}\alpha^i_n|i\rangle$;  $|i\rangle$ are the eigenstates of the plaquette and $\alpha^i_n$ are complex variational constants of the n-th plaquette which satisfy $\sum_i|\alpha^i_n|^2=1$  and determined by minimizing the energy  $E=\langle\psi|\mathcal{H}|\psi\rangle$.

Replacing Eq. (\ref{eq:psi_VA})  in Eq. (\ref{eq:Ham}) we obtain

\bea
E&=&J_0\sum_n\vec{\alpha}^{\dagger}_{n}.M_d.\vec{\alpha}_{n}-h\sum_{n,a}\vec{\alpha}^{\dagger}_{n}\cdot M^{z}_{a}\cdot\vec{\alpha}_{n}+\nn\\
&&J_1\sum_{n,\gamma,a}\left(\vec{\alpha}^{\dagger}_{n}\cdot M^{\gamma}_{a}\cdot\vec{\alpha}_{n}\right)\left(\vec{\alpha}^{\dagger}_{n+1}\cdot M^{\gamma}_{a}\cdot\vec{\alpha}_{n+1}\right)+\nn\\
&&J_2\sum_{n,\gamma,a}\left(\vec{\alpha}^{\dagger}_{n}\cdot M^{\gamma}_{a}\cdot\vec{\alpha}_{n}\right)\left(\vec{\alpha}^{\dagger}_{n+1}\cdot M^{\gamma}_{a+1}\cdot\vec{\alpha}_{n+1}\right)\nn\\
\eea
where  $M_d =$ is a diagonal matrix with elements given by the eigenvalues of the plaquette and $M^{\gamma}_{a}$ is the component  $\gamma=x,y,z$ of original spin  $\Sp_{n,a}$ of site $a$ in the basis of eigenvectors
of $\mathcal{H}_{\text{plaq}}$.

We propose that the wave function is a linear combination of the three different ground states that a single plaquette has depending on the applied magnetic field ( $|s^{(1)}\rangle$, $|t^{(1)}_1\rangle$ and $|q_1\rangle$). Although this is a minimal starting point, expected to be valid in the weak inter-plaquette regime, it however predicts the emergence of the different plateaux structures, as we show below.

The ground state is obtained using simulated annealing on lattices with PBC  and choosing the state with lowest energy per site. Simulations  were done by an exponential annealing schedule and the whole process was repeated enough times to ensure stability of results.

\begin{figure}
\includegraphics[width=1\columnwidth]{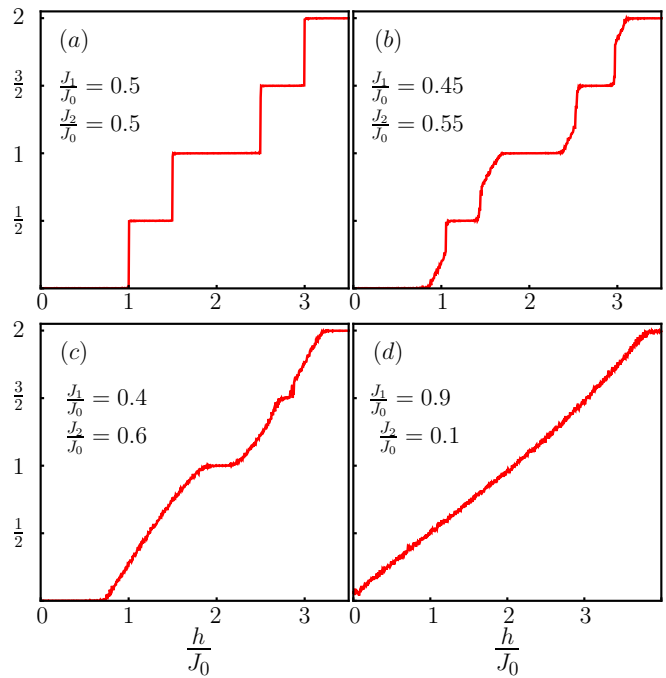}
\caption{Magnetization curves obtained from the variational approach. On the left, both interplaquette couplings are equal $J_1=J_2=0.2\,J_0$.
On the right, the couplings are slightly different $J_1=0.2\,J_0, J_2=0.17\,J_0$.
}
\label{fig:mvsh_varia}
\end{figure}

In  figure \ref{fig:mvsh_varia}  we show typical magnetization curves obtained for some values of $J_1/J_0$ and $J_2/J_0$. These $J_{1,2}/J_0$ values have been chosen  in order to illustrate the consistency of HVA with the other low energy methods and the DMRG results: that is the  emergence of different plateaux structures.
The left-upper panel shows a typical Ising-like behavior along the line of maximum frustration for the case  $J_1/J_0=J_2/J_0=0.5$.
We find a structure of plateaux separated by jumps, where there are integer plateaux at $M=0,1$ and $2$, and two half integer plateaux at $M=1/2$ and $M=3/2$. Checking the values of the $\alpha_i$ parameters,
we see that the $M=1/2$ plateau corresponds to a wave function made of the singlet state  $|s^{(1)}\rangle$ in one plaquette and the triplet  $|t^{(1)}_1\rangle$ in the other one, and $M=3/2$ correspond to  the triplet in one and a quintuplet $|q_1\rangle$ in the other one.
As we slightly move in the $J_{1,2}/J_0$ space off the diagonal $J_1=J_2$,  the effect of frustration starts to decrease.
This is reflected in the magnetization curves in a reduction of the half-integer plateaux and the change from jumps  to continuous curves  between   plateaux.
This  is shown in the right-upper panel of Fig. \ref{fig:mvsh_varia}, for the case  $J_1/J_0=0.45$ and $J_2/J_0=0.55$.
Notice that, when compared to the DMRG curve in the right-upper panel of Fig. \ref{fig:fields_hc}, the width of the plateaux is larger and the curvature between plateaux is different.
In this method, as we are only using three states, the structure of the plateaux looks more robust.
Indeed, as an example the left-lower panel of Fig. \ref{fig:mvsh_varia}   shows that for  $J_1/J_0=0.4$
and $J_2/J_0=0.6$, where no more half-integer plateaux were seen for DMRG, the $M=1/2$ plateau is not present but a small $M=3/2$ plateau survives, along  with the integer plateaux structure.
However, this method also shows that  far from the diagonal, the plateaux structure disappears completely before saturation.
This behavior is illustrated in the left-right panel of Fig. \ref{fig:fields_hc}, for   $J_1/J_0=0.9$ and $J_2/J_0=0.1$.

To summarize, the wave function proposed in the variational approach  captures qualitatively the main features of the low energy behavior of the model under a magnetic field:
when $J_1/J_2=1$ there are jumps between the different plateaux in the magnetization curve, where two of them are half integer at $M=1/2$ and $M=3/2$. As the difference between the frustrating parameters $J_1$ and $J_2$ increases, first these jumps become smoother curves and  the half integer plateaux become smaller until they disappear. As $J_1/J_2$ is further away from 1, the integer plateaux are also washed away. Therefore, although the algorithm does not guarantee convergence to the ground state we nevertheless trust that an accurate picture emerges since the results capture the main features of the low energy physics, compatible with the analytical calculations and DMRG simulations.

%

\section{Discussion}
\label{sec-Discussion-Conclusions}

In this Section we present an analysis of the extension of different phases present in the spin-tube model,
around the strong plaquette limit, focusing in the interplay between frustration and magnetic field.
The aim is to analyze the consistency of the predictions of different methods employed in the work, in particular between numerical and low energy effective models.

\begin{figure}[htb]
\includegraphics[width=0.85\columnwidth]{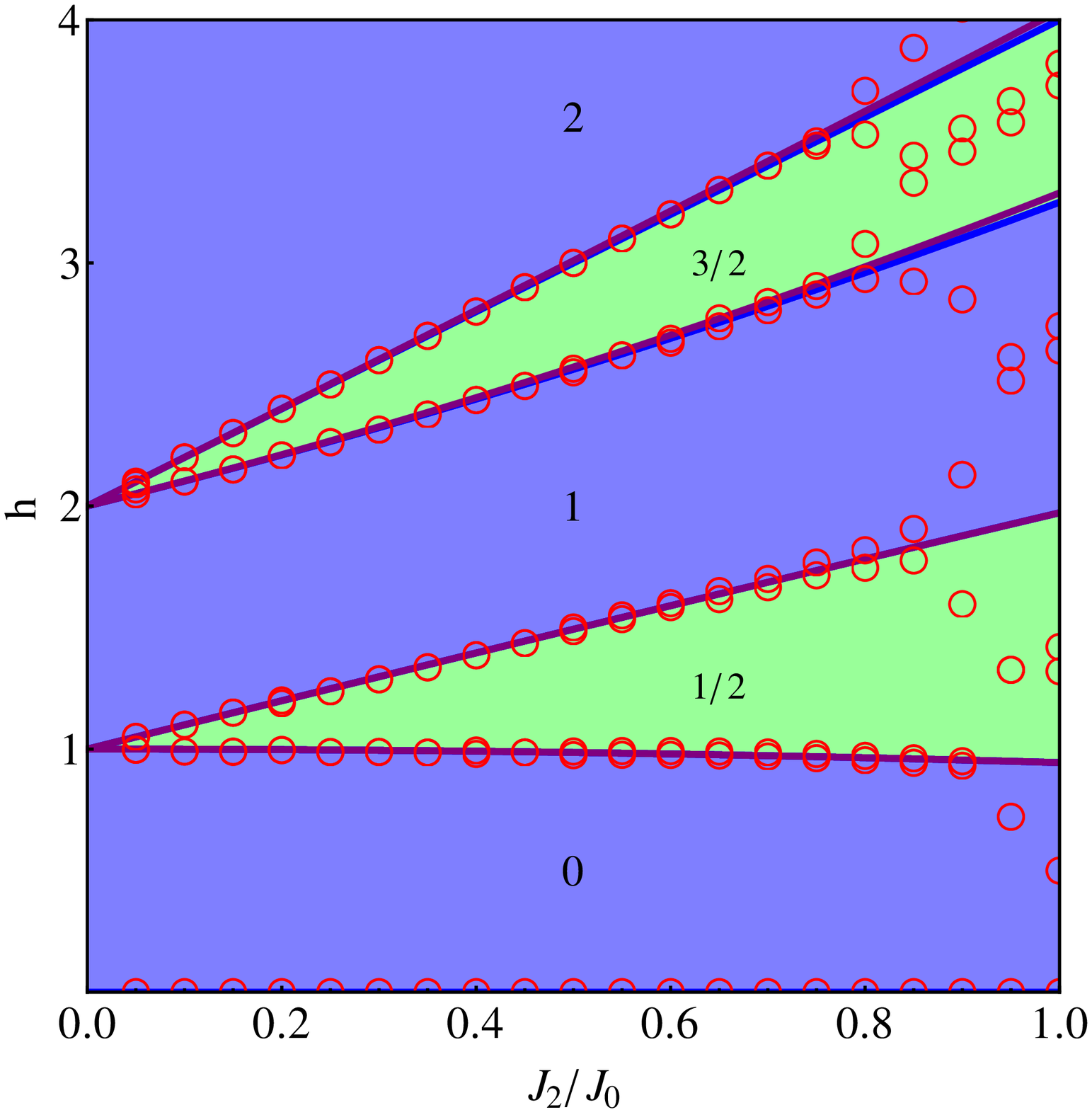}
\includegraphics[width=0.85\columnwidth]{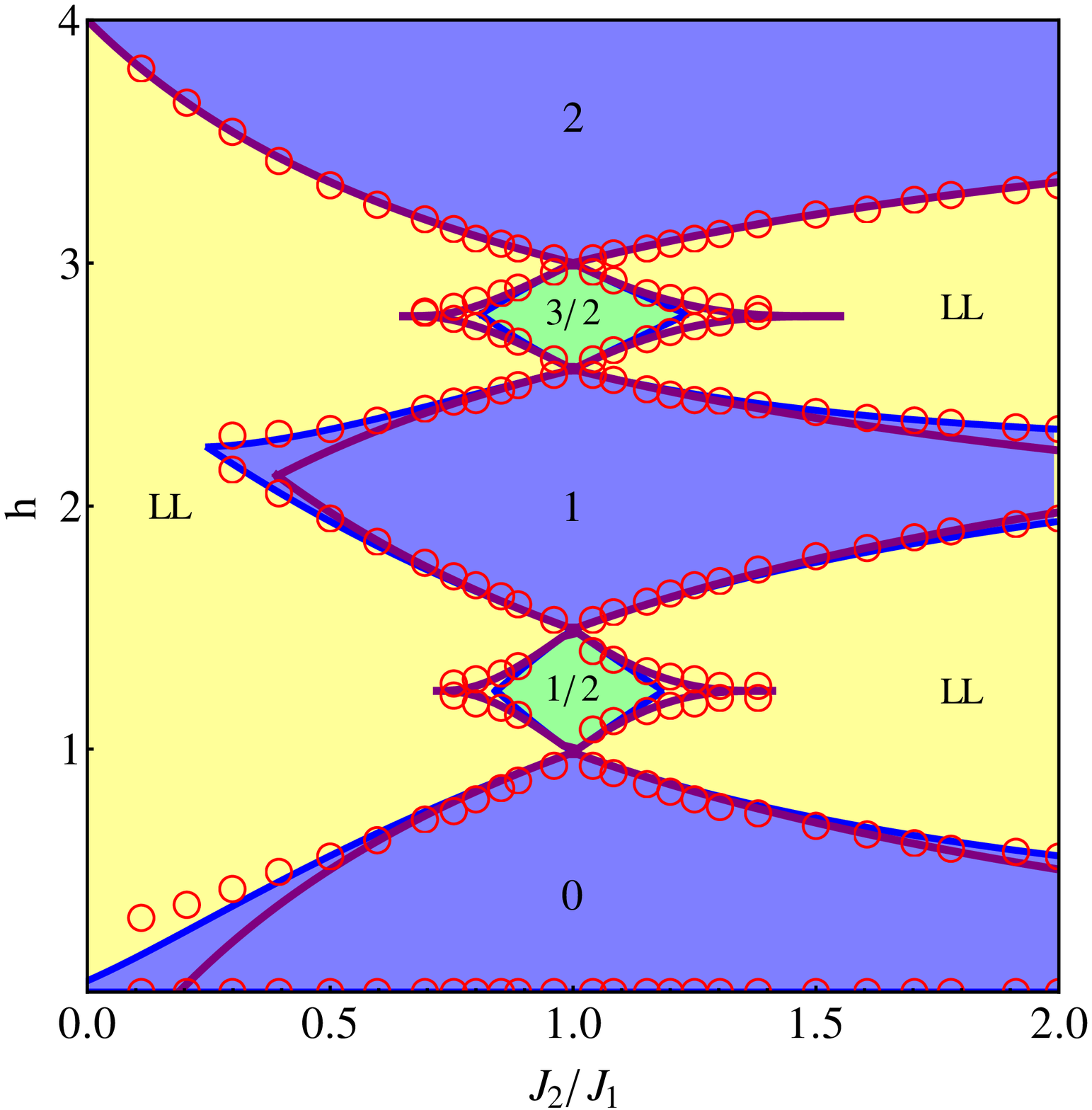}
\caption{(Color online) Top: phase diagram $h$ vs $J_2/J_0$ along the line  $J_1=J_2$ obtained by (i) the Bethe-Ansatz solution of the XXZ Hamiltonian using coefficients up to second order corrections (purple solid line); (ii) second order magnon and spinon dispersion  Eqs. (\ref{eq:whc4}), (\ref{eq:whc158}) and (\ref{eq:whc2367}) (solid blue line) and DMRG simulations for a tube of $L=20$ plaquettes (red empty circles); Bottom: phase diagram $h$ vs $J_2/J_1$  for $J_1+J_2=J_0$}
\label{fig:h_vs_beta_eff12}
\end{figure}

The main results obtained are summarized in the phase diagrams depicted in Fig. \ref{fig:h_vs_beta_eff12}. On the one hand, top panel shows magnetic phases along maximum frustration line $J_1=J_2$ vs. magnetic field $h$. Here, blue areas represent integer plateaux $M=0,1$ and $2$, whereas the green areas half-integer plateaux $M=1/2$ and $3/2$ from bottom to up, respectively.
Solid purple lines represent solutions of the second order effective Ising model, solid blue lines are solutions obtained by analyzing the closure of magnon-like dispersion of second order effective model
(Eqs.(\ref{eq:whc4}-\ref{eq:whc158})), whereas  red open circles are critical points determined numerically by means of DMRG on finite size tubes, composed by $L=20$ plaquettes with PBC and keeping $m=500$ basis states during computation.

As it can be observed, all techniques predict consistently a linear increase of plateaux width with the frustrating parameters, at least for small values of $J_2/J_0$. Second order contributions are more noticeable for $J_2/J_0 \gtrsim 0.5$, in particular for the critical lines separating
plateaux at $M=0$ and $1/2$ and $M=1$ and $3/2$.

It is important to stress that only the effective Ising model predicts strictly jumps between plateaux,  throughout the line $J_1=J_2$.

In the case of numerical DMRG predictions, the stepwise structure along $J_1=J_2$  predicted by the effective  Ising model, starts loosing validity around $J_1/J_0 \approx 0.8$. Beyond that point, along that line, deviations respect to effective model predictions are increasingly more pronounced, as shown by red circles at the right part of Fig \ref{fig:h_vs_beta_eff12} (top panel). Apart from possible finite size effects affecting numerical computations, the reason of such deviations could be intrinsic to the model. In fact, it is known that at zero magnetic field, the tube model around $J_1=J_2\approx J_0$ undergoes a first order quantum phase transition from the plaquette phase to an \a tiny spiral -like ordered phase\cite{Arlego2012}. Therefore, deviations observed around that limiting point might be an indication of the existence of such transition, even though the analysis of the effect of the magnetic field on this transition is beyond the scope of the present work. \\

On the other hand, in the lower panel of Fig. \ref{fig:h_vs_beta_eff12} we
show the extension of different plateaux structures as determined by the different techniques, as a  function of the ratio $J_2/J_1$ and magnetic field $h$, along the line $J_1+J_2=J_0$.
Note that $J_1 \leftrightarrow J_2$ symmetry is manifest in this figure, as one-half of it can be obtained from the other one. However we keep both sides (around $J_2/J_1=1$) to highlight explicitly the presence of this symmetry.

As in the upper panel of Fig. \ref{fig:h_vs_beta_eff12},
blue and green areas represent integer and half-integer plateaux, respectively. Yellow regions represent a phase where magnetization increases continuously with applied field $h$ and which we identify with Luttinger liquid-like phase within the framework of the effective model. \\
First of all note that, overall, all techniques predict two half-integer, frustration induced plateaux, which are widest on the $J_2/J_1=1$ line, and clearly tend to decrease and eventually disappear as we move further from $J_2/J_1= 1$ .\\
Also notice that integer plateaux are larger and more robust versus frustration. In fact, they even exist for isolated plaquettes, which is not the case of half-integer plateaux. The effect of coupling on integer plateaux is a renormalization, which is well captured by the low energy effective model, in the range $0.5 \leq J_2/J_1 \leq 1.5$.\\

Regarding the low-energy models results, in the lower panel of Fig. \ref{fig:h_vs_beta_eff12}, solid purple lines indicate critical fields obtained by solving the Bethe-Ansatz Eqs.(\ref{eq:Hc-BA}). Although $O(J_1,J_2)$ results bring a good description, in order to improve these estimates around $J_1=J_2$, we have retained the XXZ model parameters up to terms linear in $(J_1-J_2)$.
The procedure to obtain these critical fields consists in replacing  $J_{xy}$, $J_{zz}$ and $h^{\text{eff}}$ of Eqs.(\ref{eq:LEH-constants-1},\ref{eq:LEH-constants-2}) into Eqs.(\ref{eq:Hc-BA}), retaining $O(J_1-J_2)$ terms, and solving numerically Eqs.(\ref{eq:Hc-BA}) for $h$ vs $J_2/J_1$, $(J_1+J_2=J_0)$.
This gives rise to the critical lines in lower and upper half of Fig. \ref{fig:h_vs_beta_eff12} (lower panel), corresponding to Eqs.(\ref{eq:LEH-constants-1}) and Eqs.(\ref{eq:LEH-constants-2}), respectively.
In particular curves bordering integer plateaux are determined by the first pair $(\pm)$ of Eqs.(\ref{eq:Hc-BA}), whereas half-integer plateaux by the second pair of Eqs.(\ref{eq:Hc-BA}).
Note that this procedure is valid, since the other terms, $K_{xy}$ and $J_{xyz}$, in effective model of Eqs.(\ref{eq:LEH-constants-1}) and Eqs.(\ref{eq:LEH-constants-2}) are of $O((J_1-J_2)^{2})$.

Let us now compare Bethe-Ansatz results with critical lines obtained by analyzing the closure of magnon and spinon-like dispersions, given by Eqs.(\ref{eq:whc4}-\ref{eq:whc2367}), for integer and half-integer plateaux, obtained from the effective  $O(J_1^{2},J_2^{2})$ Hamiltonian in Eq. (\ref{eq:Heff_LEH}). These results are shown with bold blue lines in the lower panel of Fig. \ref{fig:h_vs_beta_eff12}. Note that although both Bethe-Ansatz and dispersion calculation
are in very good agreement, around the line $J_1=J_2$, dispersion analysis predicts a smaller  half-integer plateaux and, more important, tends to round the critical line at the end points. This could be due to the fact that Bethe-Ansatz, even though it is constructed
on a perturbative model, provides a non-perturbative solution, which is able to predict singularities. In contrast, any finite order perturbative dispersion calculation will be unable to reproduce the singular shape at the end points.

For the same reason it is not expected that the variational approach will be quantitatively precise in the determination of critical lines. In fact, the variational method (not shown in this panel), although it predicts qualitatively well the presence of half-integer plateaux, overestimates its range of existence, and also tends to round the critical line at the end points. \\
Finally, DMRG technique, although it provides results which are susceptible to finite size effects,
has the advantage that it is not perturbative and does not depend
on the adiabatic connection with the phase of isolated plaquettes, as the other methods.
In particular, it is able to describe more precisely the non-analyticity of ending points of half-integer plateaux,
as shown with red open circles in the lower panel of Fig. \ref{fig:h_vs_beta_eff12}, for $L=20$ and $m=500$.
It is interesting to note that dispersion calculation is in  better agreement with DMRG results, as compared with Bethe-Ansatz,
regarding integer-plateaux far from the $J_1=J_2$ line.\\

\section{Conclusions}
\label{sec-Conclusions}
%
In conclusion, we have studied quantum phases of a frustrated spin-1/2 four-leg tube in an external magnetic field, around the isolated plaquette  limit, by means of low energy perturbative and variational methods, complemented with numerical DMRG simulations.\\
We observe that frustrating inter-plaquette couplings induce the emergence of half-integer plateaux in the magnetization curves, as well as a renormalization of integer plateaux, already present in the case of decoupled plaquettes.\\
Low energy effective models capture the essential features of the system, and provide physical insight about the nature of the different phases present in the system.\\ On the other hand, DMRG numerical simulations allowed us to check the range of validity of the effective models around the plaquette phase.\\
Finally, we would like to mention that the exploration of other regions of parameter space of the model, beyond plaquette phase, which have not been considered here, remains as an open issue. In particular, the analysis around the spiral phase of the model is an interesting topic that clearly deserves future investigations.

\section*{Acknowledgments}

The authors specially thank Pierre Pujol  and Gerardo L. Rossini for fruitful discussions. This work
was partially supported by CONICET (PIP 0747) and
ANPCyT (PICT 2012-1724).


\end{document}